\documentclass [
superscriptaddress,
reprint,
 amsmath,amssymb,
 showkeys
]{revtex4-1}
\setcitestyle{super}
\usepackage{graphicx}
\usepackage{dcolumn}
\usepackage{bm}
\usepackage{color}
\usepackage{float}
\usepackage[utf8]{inputenc}
\usepackage[mathlines]{lineno}
\usepackage{multirow}
\usepackage{hyperref}
\usepackage{amsmath}
\usepackage{xr}
\usepackage{soul,xcolor}
\soulregister\cite7
\soulregister\ref7
\setstcolor{blue}
\newcommand{\wn}{cm$^{-1}$}

\newcommand{\doHMN}[2]{%
  \begingroup\lccode`~=`#1
  \lowercase{\endgroup\let~}#2%
  \mathcode`#1="8000
}

\begin{document}

\title{Anharmonic Lattice Vibrations in Small-Molecule Organic Semiconductors}

\author{Maor Asher}
\affiliation{Department of Materials and Interfaces, Weizmann Institute of Science, Rehovot 76100, Israel}
\author{Daniel Angerer}
\affiliation{Institute of Theoretical Physics, University of Regensburg, 93040 Regensburg, Germany}
\author{Roman Korobko}
\affiliation{Department of Materials and Interfaces, Weizmann Institute of Science, Rehovot 76100, Israel}
\author{Yael Diskin-Posner}
\affiliation{Chemical Research Support, Weizmann Institute of Science, 234 Herzl Street, Rehovot 76100, Israel}
\author{David A. Egger}
\email{david.egger@tum.de}
\affiliation{Department of Physics, Technical University of Munich, 85748 Garching, Germany}
\author{Omer Yaffe}
\email{omer.yaffe@weizmann.ac.il}
\affiliation{Department of Materials and Interfaces, Weizmann Institute of Science, Rehovot 76100, Israel}

\date{\today}

\begin{abstract}

The intermolecular lattice vibrations in small-molecule organic semiconductors have a strong impact on their functional properties. 
Existing models treat the lattice vibrations within the harmonic approximation.
In this work, we use polarization-orientation (PO) Raman measurements to monitor the temperature-evolution of the symmetry of lattice vibrations in anthracene and pentacene single crystals.
Combined with first-principles calculations, we show that at 10~K the lattice dynamics of the crystals are indeed harmonic.
However, as the temperature is increased specific lattice modes gradually lose their PO dependence and become more liquid-like. 
This finding is indicative of a dynamic symmetry breaking of the crystal structure and shows clear evidence of the strongly anharmonic nature of these vibrations.
Pentacene also shows an apparent phase transition between $80-150$~K, indicated by a change in the vibrational symmetry of one of the lattice modes.
Our findings lay the groundwork for accurate predictions of the electronic properties of high-mobility organic semiconductors at room temperature.

\end{abstract}

\maketitle

Small-molecule semiconducting crystals are studied extensively due to their potential for (opto)electronic applications such as light-emitting diodes, field-effect transistors, and solar cells.\cite{Horowitz1999,ThejoKalyani2012,Cheng2009,Ohno2005, Coropceanu, Podzorov2013, Karl, brutting2006physics}
Linear oligoacenes are an archetypical family of such organic crystals that serves as an excellent testbed to study intrinsic properties of $\pi$-conjugated solids.
This is due to their well-defined crystal structure~\cite{Campbell1962,Simpkins1998,Anthony2008} and their wide range of tunable optical and electronic properties.\cite{SaNchez-Carrera,Lipari1975,Deng,Hummer}

Thermal fluctuations of the nuclei in organic crystals have a significant role in determining their optical~\cite{Hestand2015}, electronic~\cite{Illig2016} and thermal~\cite{Coleman1973} properties.
These fluctuations stem from both intermolecular and intramolecular vibrations, which contribute to the dynamic disorder.\cite{Fratini2017,Illig2016,Wang2010a}
Contemporary theoretical studies are in agreement that low-frequency ($<$150~\wn) lattice vibrations dominate charge carrier mobility in organic crystals via nonlocal electron-phonon interactions.~\cite{Troisi2006,Fratini2017,Wang2007,Fratini2016,Schweicher2019}

To the best of our knowledge, current models for charge transport and optical properties of organic semiconductors use the harmonic approximation to describe lattice vibrations.\cite{Wang2011,Wang2013,Coropceanu,Troisi2011a, Fratini2016,Fetherolf2019}  
The anharmonic components of the lattice vibrations, i.e., phonon-phonon interactions \footnote{Formally, anharmonic structural dynamics is defined by the higher-order ($n>2$) terms of the Taylor expansion for the crystal potential energy in terms of nuclear displacements from their equilibrium position.~\cite{Dove2003} } are therefore entirely neglected.
However, organic crystals are known to be mechanically soft~\cite{Vaidya1971}, exhibit large molecular displacements ~\cite{Sosorev2019} and have large thermal expansion coefficients~\cite{Haas2007} all of which are indicative of strong lattice anharmonicity.\cite{Andritsos2013,Dove2003}
Strongly anharmonic lattice dynamics are expected to have a disruptive effect on the electronic coupling between the molecules, and therefore they may profoundly impact charge transport in the organic crystal.
Therefore, it is imperative to characterize the degree of lattice anharmonicity in small-molecule organic crystals.
The main experimental methods for measuring the lattice dynamics to infer about anharmonicity are THz spectroscopy, inelastic neutron scattering and Raman spectroscopy.~\cite{Schweicher2019}

\begin{figure*}
\includegraphics[scale=0.123]{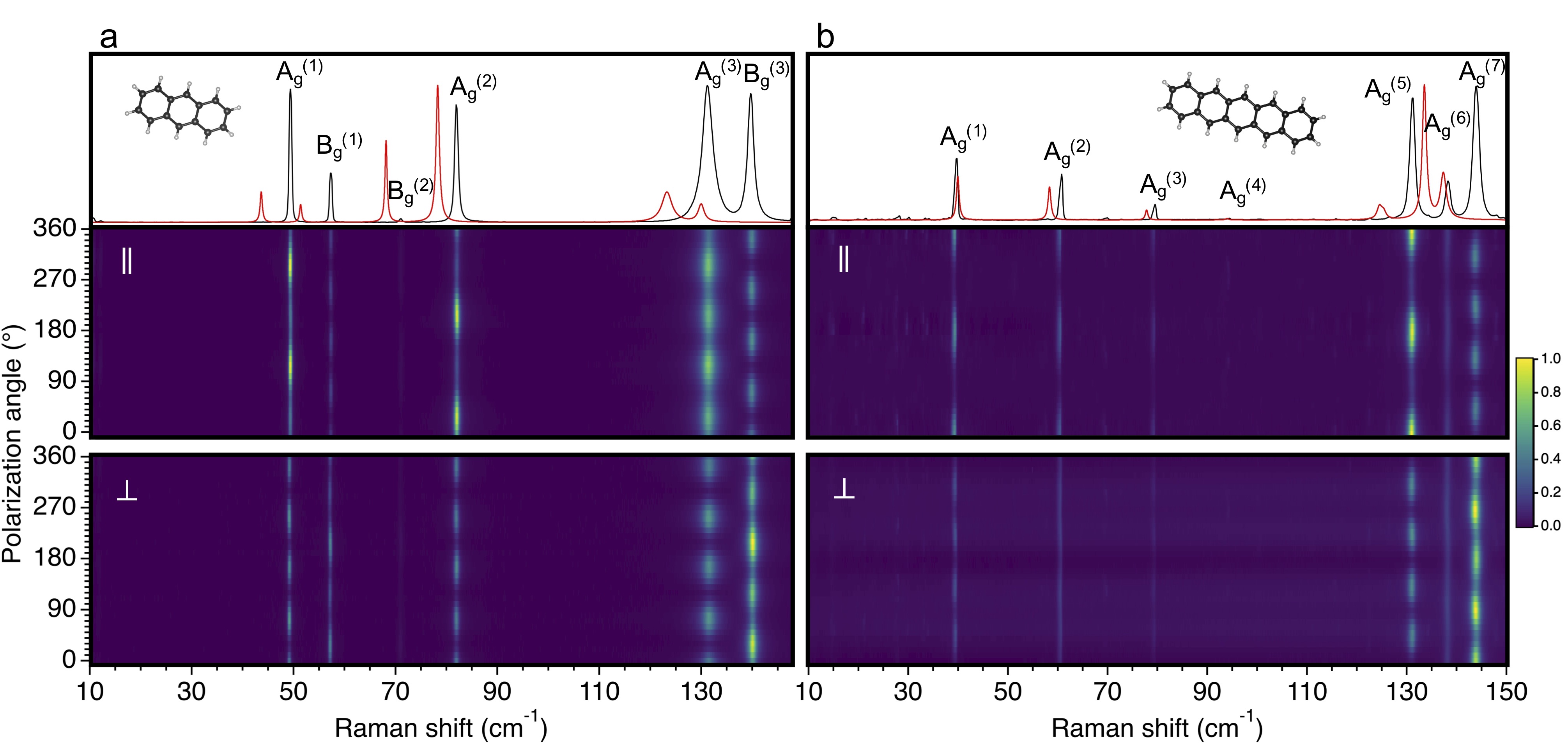}
\caption{\label{fig:PO_image}Unpolarized Raman spectra (top panel) and their polarization-dependence in parallel (mid panel) and perpendicular configuration (lower panel) at 10~K for (a) anthracene and (b) pentacene (001) single crystals. The unpolarized experimental spectra are shown in black and the DFT-calculated ones in red.}
\end{figure*}

In this study, we use polarization-orientation (PO) Raman spectroscopy to directly probe the evolution of \textit{phonon symmetry} with temperature.
We present unambiguous experimental evidence for strongly anharmonic behavior of specific lattice vibrations in oligoacenes crystals.
This anharmonic behavior is expressed by temperature-activated symmetry breaking of the average crystal structure well-below their melting temperature. 
By combining PO Raman with first-principles calculations based on density functional theory (DFT), we show how specific lattice modes gradually lose their vibrational symmetry and become liquid-like due to enhancement of vibrational anharmonicity as the temperature is increased.

Single crystals of anthracene and pentacene were prepared via physical vapor transport (PVT).~\cite{Simpkins1998}
We measured their crystal structure and orientation by conducting single-crystal x-ray diffraction (XRD) measurements (see SI, Figure S1).
The crystal orientation for both was found to be (001), in-line with previous reports.~\cite{Mattheus2003a, Zeng2007}

Figure~\ref{fig:PO_image} shows the raw PO Raman data for the single crystals of anthracene and pentacene at 10~K.
In this measurement, the crystal oriented along the crystallographic $ab$ plane is excited by a linearly-polarized laser (785~nm).
The scattered light is then filtered by an analyzer for polarization parallel and perpendicular to the incident light.
This measurement is repeated after rotating the polarization of the incident light while the sample position is fixed.
The false-color plots show the fluctuations in scattering intensity as a function of the angle between the polarization of the incident light and an arbitrary axis in the $ab$ plane.
In the top panel of Figure~\ref{fig:PO_image}, we present the normalized unpolarized (i.e. integrated over all polarization angles) Raman spectra (black line) along with the DFT-calculated Raman intensities (red line). 
The agreement between theory and experiment is overall very good (i.e., frequencies differ by at most 10~cm$^{-1}$).
We note that the apparent redshifts (e.g., in the higher-energy part of the pentacene spectrum) can be explained by a tendency of underbinding~\cite{Ambrosetti2014} in organic crystals when the here-applied, many-body dispersion (MBD) correction is used.\cite{Tkatchenko2012}
In the SI, section S2, we demonstrate the superior performance of the MBD approach compared to using the regular Tkatchenko-Scheffler (TS) method.\cite{Tkatchenko2009,Kronik2014,Hermann2017,Bedoya-Martinez2018} 

To analyze the PO results of Figure~\ref{fig:PO_image}, we first fit each spectrum to the product of the Bose-Einstein distribution and a multi-Lorentzian line shape (see SI, section S3).
Then, we extract the integrated intensity of the deconvolved Lorentzian of each peak and monitor its fluctuation with the polarization angle.
Finally, we fit the intensity fluctuations for each mode using a model proposed by Kranert et al.~\cite{Kranert2016} in order to extract the Raman tensors (see SI, section S4).
This model considers the anisotropic nature of the crystals (i.e. birefringence), the scattering cross-section dependence of each mode on the vibration frequency, and the coordinates of the crystal with respect to the optical table. 

We perform a factor group analysis to extract the number of lattice modes and the form of the Raman tensors.~\cite{Laboratories1981}
Anthracene has a monoclinic crystal structure with a $P2_{1}/a$ space group for which factor group analysis predicts 6 Raman-active lattice modes, namely 3 $A_{g}$ modes and 3 $B_{g}$ modes.
The form of the Raman tensors is,

\begin{equation}
 R_{A_{g}}=\begin{pmatrix} a&0&e\\0&b&0\\e&0&c \end{pmatrix} ~,~  R_{B_{g}}=\begin{pmatrix} 0&d&0\\d&0&f\\0&f&0 \end{pmatrix}
\label{eq:Raman_tensor}
\end{equation}
All 6 modes are observed in the data presented in Figure~\ref{fig:PO_image}a.
 
Pentacene has a triclinic crystal structure with a $P\bar{1}$ space group for which factor group analysis predicts 6 Raman-active lattice modes.
All of them have $A_{g}$ symmetry with a Raman tensor form,

\begin{equation}
 R_{A_{g}}=\begin{pmatrix} a&d&e\\d&b&f\\e&f&c \end{pmatrix}
\label{eq:Raman_tensor1}
\end{equation}
The data presented in Figure~\ref{fig:PO_image}b shows 7 modes. 
The additional mode is due to the mixing of inter- and intramolecular vibrations, as discussed previously.~\cite{Filippini1984}

\begin{figure}
\centering
\includegraphics[scale=0.125]{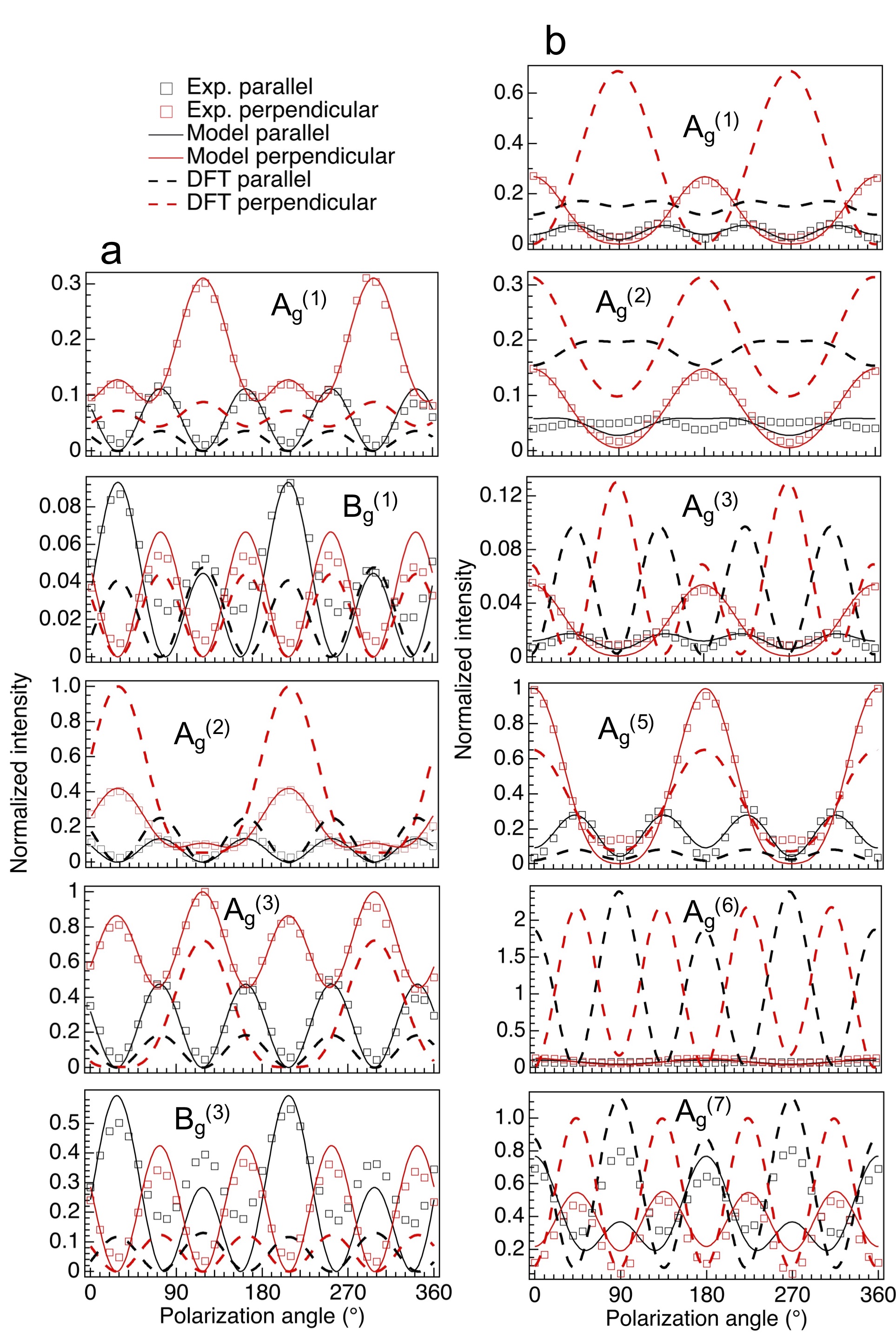}
\caption{\label{fig:PO}PO mode intensity for the low-frequency modes (see Figure~\ref{fig:PO_image}) of (a) anthracene and (b) pentacene. We compare the results from experiment (squares), theoretical model (solid lines) and  DFT-calculated Raman tensor elements (dashed line) for parallel (red) and perpendicular (black) polarization configurations.}
\end{figure}

To extract the Raman tensor of each mode, we perform a global fit (i.e. simultaneous fit to all peaks) to the PO dependence of the integrated intensity (see SI, section S5).
\footnote{Due to the (001) orientation of the crystals and backscattering geometry, the extract Raman tensors include only the $x$ and $y$ components.}
\footnote{The signal from the $B_{g}^{(2)}$ mode of anthracene and the $A_{g}^{(4)}$ of pentacene is very weak and their polarization dependence can be hardly resolved. This is probably due to the low values of their Raman tensor components. Therefore, these modes are not included in the analysis.}
For each mode in Figure~\ref{fig:PO_image} we present the fit results in Figure~\ref{fig:PO} (solid lines) along with the deconvolved integrated intensity (squares) and results of the same model based on the DFT-calculated Raman tensors (dashed lines).
We find an overall good agreement between the experimental data, the global fit, and the DFT results.
This shows that at 10~K the lattice dynamics are well-captured by the harmonic approximation that is inherent to our DFT-based Raman calculations. 
Interestingly, the agreement is somewhat less favorable for pentacene than anthracene, especially in regard to the $A_{g}^{(5)}$ mode in pentacene, which will be discussed below. 

\begin{figure}
\centering
\includegraphics[scale=0.125]{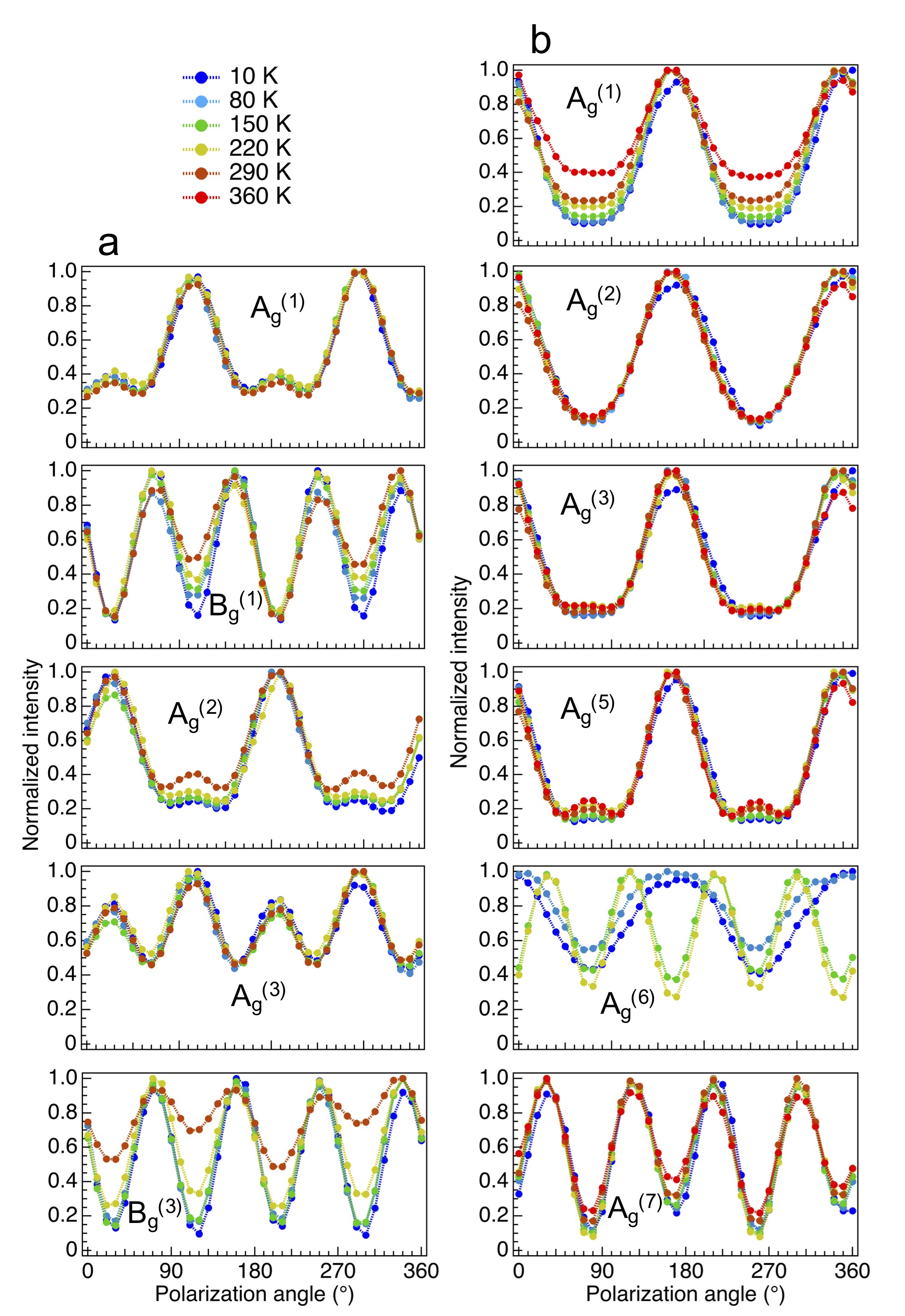}
\caption{\label{fig:PO_temp}Experimentally-measured temperature-dependence of the polarization-dependence for the six lowest-frequency modes of (a) anthracene and (b) pentacene in parallel configuration. The intensities were normalized for each mode at each temperature.}
\end{figure}

Figure~\ref{fig:PO_temp} shows the temperature-dependent (10-360~K) PO data in parallel configuration for anthracene (a) and pentacene (b).
The experimental procedure and analysis are equivalent to those described for the data in Figure~\ref{fig:PO} and the raw Raman spectra for each temperature are presented in section S6 in the SI.
For a perfectly harmonic crystal and in the absence of structural phase transitions, the PO behavior is expected to be temperature independent.
Remarkably, we find that $B_{g}^{(1)}$ and  $B_{g}^{(3)}$ modes in anthracene and  $A_{g}^{(1)}$ mode in pentacene gradually lose their PO dependence as temperature is increased.\footnote{This temperature range is well below the crystals melting temperature which is 212~C$^\circ$ for anthracene,~\cite{Chaiken1966} pentacene decomposes above 600~C$^\circ$~\cite{Fulem2008}).}
As the PO dependence becomes weaker, these modes become more liquid-like, since for liquids the Raman intensity is PO independent (see SI, section S8 for an example of a PO measurement of a liquid).
This is despite the fact that single crystal XRD at 100~K and 293~K show no change in the average crystal structure (see SI, section S1). 
We note that the same trend is observed also in the PO of the perpendicular configuration (see SI, section S7).

Since the PO of each mode reflects its symmetry, our findings reveal a continuous and mode-specific symmetry breaking in anthracene and pentacene crystals. 
This symmetry breaking must stem from the temperature-driven lattice fluctuations in the crystals since the trends in the PO intensities are fully reversible with temperature, which rules out the appearance of static disorder.
 
We interpret this dynamic symmetry breaking as a manifestation of vibrational anharmonicity that increases with temperature. Traditional expressions of anharmonicity include phonon 'softening' (decrease in vibrational energy) and decreasing phonon lifetimes (increase in vibrational linewidth).~\cite{Lucazeau2003}
The observed symmetry breaking \textit{is not accompanied by any anomalous effect in the temperature dependence of the modes energies or their linewidth} (see SI, section S9).
This is consistent with DFT-calculated phonon density of states of anthracene and pentacene (see SI, section S10) that show no distinct features that would imply a strong coupling of the anomalous vibrations (i.e. the vibrations that exhibit symmetry breaking) to the thermal bath (i.e. acoustic phonons).
Importantly, strong anharmonicity does not necessarily lead to mode softening and decreasing phonon lifetimes. Anharmonic effects are usually modeled by three- and four-phonon decay processes ~\cite{Lucazeau2003,Ouillon1984,Lan2012} where a dominance of four-phonon processes can lead to phonon hardening and increasing lifetimes.~\cite{Ravindran2003,Ponosov2004,Li2011} 
A possible explanation of this  finding is that due to anharmonic effects, the phonon dispersion relation evolves with temperature.~\cite{Hellman2013}

A second important finding present in Figure~\ref{fig:PO_temp}b is that the $A_{g}^{(6)}$ mode of pentacene changes its PO periodicity from 180$^\circ$ below 80~K to 90$^\circ$ above 150~K. This change is reversible with temperature and indicative of a profound change in the Raman tensor elements of this vibration.
Unlike the gradual PO loss discussed in the previous paragraph, this change in PO periodicity is accompanied by changes in the temperature evolution of the unpolarized Raman spectra, since the temperature dependence of the $A_{g}^{(6)}$ mode width shows a subtle change in its slope between 80-150~K (see SI, section S9).
These observations may be indicative of a phase transition (which is an explicit anharmonic phenomenon) to a different polymorph or could be another manifestation of the complex effect of anharmonicity on the lattice vibrations of the crystal.
Yet, we are unaware of a reported phase transition in pentacene in this temperature range.~\cite{Brillante2002,Farina2003a,Mattheus2003,Mattheus2003a}
It is at first sight surprising that such a large change in PO behaviour has a relatively small effect on the unpolarized Raman spectrum. 
However, phase transitions between polymorphs can be very subtle and have little~\cite{Brillante2002} or no effect~\cite{Guo2017} on the spectrum because Raman scattering probes only Raman-active modes at the center of the Brillouin zone.

In conclusion, we discover a new manifestation of anharmonic vibrational behavior in anthracene and pentacene single crystals, by using temperature-dependent Raman PO measurements and first-principles calculations.
We detect a liquid-like behavior in specific lattice modes, indicated by a gradual loss of the PO dependence at temperatures that are well below the melting point of the crystals.
This PO-dependence loss is a signature of dynamic symmetry breaking of the crystal structure due to anharmonic thermal fluctuations.
The mechanism that leads to mode-specific symmetry breaking is still unclear and requires further study.
We also detect what seems to be a subtle phase transition in pentacene between 80-150~K, manifested in a change in vibrational symmetry of the $A_{g}^{(6)}$ mode.
Since the low-frequency modes are known to be important sources of electron-phonon interactions in organic crystals, our findings have many interesting implications, since they question the validity of the harmonic approximation when describing the transport properties of these materials. 

\subsection*{Experimental Section} 

\textit{Crystal Growth:} Single crystals of anthracene and pentacene were grown by physical vapor transport (PVT). For anthracene, a 99$\%$  powder (Sigma-Aldrich), a temperature of 135~C in the tube furnace and an argon flow of 60~ml~min$^{-1}$ were used. For pentacene, a 95$\%$ powder (Toronto Research Chemicals) was used with a similar method except for the temperature of the tube furnace being 280~C. The temperatures of the furnace were set to lower temperatures from those in the literature~\cite{Simpkins1998} to ensure the growth of high-quality single crystals. Crystals were also grown by using a two-zone furnace with the same source temperature and deposition temperature of 100~C and 200~C for anthracene and pentacene respectively. The PO Raman results of these crystals were very similar, suggesting similar crystal quality.

\textit{Temperature Dependent PO Raman:} A custom-built dispersive Raman spectrometer was used to conduct the Raman measurements. The system is based on a 1~m long Horiba FHR-1000. Notch filters are included in the system to allow access to the low-frequency region ($>$10~cm$^{-1}$) and simultaneous acquisition of the Stokes and anti-Stokes signal. To avoid photo-luminescence (PL) a 785~nm Toptica diode laser was used. For pentacene, a small PL signal was detected at some temperatures so the laser intensity on the sample was kept below 2~mW to avoid sample heating. No such problem arose in the case of anthracene so the laser intensity on the sample was kept at approximately 30~mW. The system included also a 50x objective and a 1800~mm$^{-1}$ grating. The spectral resolution was approximately 0.15~cm$^{-1}$. To control the polarization of the incident and scattered light, half-wave plates and a polarizer-analyzer combination was used (see SI, section S11). The temperature was set and controlled by a Janis cryostat ST-500 and a temperature controller by Lakeshore model 335.
The polarization dependence of anthracene and pentacene (steps of 10$^{\circ}$) was measured in parallel and perpendicular configurations at 10$~$K, 80$~$K, 150$~$K, 220$~$K, 290$~$K and 360$~$K. For anthracene, no measurement at 360$~$K was performed due to the sublimation of the sample.

\textit{First-Principles Calculations:} DFT calculations were performed with the projector augmented-wave method~\cite{Joubert1999} as implemented in VASP.\cite{Kresse1996a}
Exchange-correlation was described using the PBE functional.\cite{Perdew1996} 
Dispersive corrections were computed using the MBD method \cite{Tkatchenko2012,Ambrosetti2014,Bucko2013,Bucko2016}
Unless noted otherwise, a plane-wave cutoff energy of 900~eV was used, and $3\times 4\times 3$ and $4\times 3\times 2$ $\Gamma$-centered k-point grids were applied for anthracene and pentacene, respectively. 
The unit cells of the crystals were optimized in internal coordinates with Gadget~\cite{Bucko2005}, applying a force threshold of 10 meV/\mbox{\normalfont\AA}. Phonon frequencies were calculated with the phonopy package~\cite{Togo2015}, using the finite-displacement method and a plane-wave cutoff energy of 800~eV.
Raman tensors associated with the respective vibrational modes were obtained using the phonopy-spectroscopy package.\cite{Skelton2017} In the experiment, we do not measure the $z$ component of the Raman tensor because we are limited to a specific crystal orientation (see above). Therefore, the unpolarized Raman intensities were calculated from the Raman tensors using only the $2\times 2$ submatrices of the tensors that are associated with their $x$ and $y$ components. In these calculations, we added the parallel and perpendicular intensities, integrated over the polarization angle, and used the experimentally-determined linewidth for each mode in a Lorentzian broadening.

\textit{X-ray Crystallography:} Single-crystal XRD measurement for pentacene was performed using Rikagu XtaLab Pro dual-source diffractometer equipped with PILATUS 200 detector and microfocus and CuK$\alpha$ radiation. For anthracene, the measurement was performed using Bruker APEX-II diffractometer equipped with KappaCCD detector and microfocus and MoK$\alpha$ radiation. The measurements were taken at 100~K after cooling at a rate of 1~K~min$^{-1}$, and at 293~K. 

\subsection*{Supporting Information} 
Supporting Information is available from the Wiley Online Library or from the author.

\subsection*{Conflict of Interest}

The authors declare no conflict of interest.

\subsection*{\label{Acknowledgments}Acknowledgements}
We thank Tsachi Livneh for fruitful discussions and Lior Segev for software development.
OY acknowledges funding from: BSF (grant No. 2016362), ERC (850041 - ANHARMONIC),  Benoziyo EF, Ilse Katz Institute, H. C. Krenter Institute, Soref Fund, C. Stiftung, A. \& S. Rochlin Foundation, E. A. Drake and R. Drake and the Harold Perlman Family.  DAE acknowledges  funding from: Alexander von Humboldt Foundation within the framework of the Sofja Kovalevskaja Award, the Technical University of Munich - Institute for Advanced Study (grant Agreement No. 291763) and by the Deutsche Forschungsgemeinschaft  (EXC 2089/1 - 390776260).

\end{document}


\subsection*{Supporting Information} 

\noindent\large{\textbf{Anharmonic Lattice Vibrations in Small-Molecule Organic Semiconductors}}\\
\noindent\large{\textit{Maor Asher, Daniel Angerer, Yael Diskin-Posner, Roman Korobko, David Egger, and Omer Yaffe$^{\ast}$ }} \\

\renewcommand{\thepage}{S\arabic{page}}  
\renewcommand{\thesection}{S\arabic{section}}   
\renewcommand{\thetable}{S\arabic{table}}   
\renewcommand{\thefigure}{S\arabic{figure}}
 \setcounter{page}{1}

\section{\label{xrd}X-ray diffraction measurements}

As mentioned in the main test, we perform single-crystal XRD measurements on anthracene and pentacene to confirm their crystal structure and orientation at 100~K and 293~K. The results of these measurements are shown in the cif files attached to the SI and summarized in \ref{tbl:XRD_anthracene} and \ref{tbl:XRD_pentacene}.\\

\begin{table}[ht]
\large
  \caption{Single crystal XRD results of anthracene.}
\center
  \label{tbl:XRD_anthracene}
  \begin{tabular*}{0.9\textwidth}{@{\extracolsep{\fill}}llll}
    \hline
    Temperature (K) & 100 & 293 \\
    \hline
    CCDC & 1947046 & 1965794 \\
    Diffractometer & \multicolumn{2}{c}{Burker APEX 2} \\
    Formula &  \multicolumn{2}{c}{C$_{14}$H$_{10}$}\\
    Formula weight & \multicolumn{2}{c}{178.22} \\
    Crystal system & \multicolumn{2}{c}{Monoclinic} \\
    Space group & \multicolumn{2}{c}{$P~2_{1}/c$} \\
    Crystal size(mm) & \multicolumn{2}{c}{0.500x0.400x0.050} \\
    Crystal color and shape & \multicolumn{2}{c}{colorless plate} \\
    Wavelength (\mbox{\normalfont\AA}) & \multicolumn{2}{c}{0.71073} \\
    a (\mbox{\normalfont\AA}) & 9.2877(8) & 9.452(2) \\
    b (\mbox{\normalfont\AA}) & 5.9902(5) & 6.0074(13) \\
    c (\mbox{\normalfont\AA}) & 8.4127(8) & 8.5421(18) \\
    $\alpha$ ($^{\circ}$) & 90 & 90 \\
    $\beta$ ($^{\circ}$) & 102.526(4) & 103.512(4) \\
    $\gamma$ ($^{\circ}$) & 90 & 90 \\
    Volume (\mbox{\normalfont\AA}$^{3}$) & 456.90(7) & 471.62(17) \\
    Z & \multicolumn{2}{c}{2} \\
    $\rho _{calculated}$ (g cm$^{-1}$) & 1.295 & 1.255 \\
    $\mu$ (mm$^{-1}$) & 0.073 & 0.071 \\
    No. of reflection (unique) & 27441(3852) & 11292(2280) \\
    R$_{int}$ & 0.0884 & 0.0732 \\
    Completeness to $\theta$ (\%) & 98.4 & 99.8 \\
    Data / restraints / parameters & 3852 / 0 / 64 & 2280 / 0 / 64 \\
    Goodness-of-fit on  $F^{2}$ & 1.032 & 1.094 \\
    Final $R_{1}$ and $wR_{2}$ indices [I $>$ 2$\sigma$(I)] & 0.0524, 0.1411 & 0.0842, 0.2508 \\
    $R_{1}$ and $wR_{2}$ indices (all data) & 0.0818, 0.1601 & 0.1349, 0.3094 \\
    Largest diff. peak and hole (e \mbox{\normalfont\AA}$^{3}$) & 0.411 and -0.467 & 0.491 and -0.383 \\
	
    \hline
  \end{tabular*}
\end{table}

\begin{table}[ht]
\large
  \caption{Single crystal XRD results of pentacene.}
\center
  \label{tbl:XRD_pentacene}
  \begin{tabular*}{0.9\textwidth}{@{\extracolsep{\fill}}llll}
    \hline
    Temperature (K) & 100 & 293 \\
    \hline
    CCDC & 1947047 & 1965795 \\
    Diffractometer & \multicolumn{2}{c}{Rigaku Xtalab PRO}  \\
    Formula & \multicolumn{2}{c}{C$_{22}$H$_{14}$} \\
    Formula weight & \multicolumn{2}{c}{278.33}  \\
    Crystal system & \multicolumn{2}{c}{Triclinic} \\
    Space group & \multicolumn{2}{c}{$P\bar{1}$}  \\
    Crystal size(mm) & \multicolumn{2}{c}{0.254x0.081x0.013} \\
    Crystal color and shape & \multicolumn{2}{c}{Dark-blue plate} \\
    Wavelength (\mbox{\normalfont\AA}) &  \multicolumn{2}{c}{1.54184} \\
    a (\mbox{\normalfont\AA}) &  6.2746(2) & 6.26910(10) \\
    b (\mbox{\normalfont\AA}) &  7.6567(2) & 7.7769(2) \\
    c (\mbox{\normalfont\AA}) & 14.3616(6) & 14.5343(5) \\
    $\alpha$ ($^{\circ}$) & 76.938(3) &  76.464(2) \\
    $\beta$ ($^{\circ}$) & 88.242(3) & 87.679(2) \\
    $\gamma$ ($^{\circ}$) & 84.325(2) & 84.692(2) \\
    Volume (\mbox{\normalfont\AA}$^{3}$) & 668.80(4) & 685.84(3) \\
    Z &   \multicolumn{2}{c}{2}  \\
    $\rho _{calculated}$ (g cm$^{-1}$) & 1.382 & 1.348\\
    $\mu$ (mm$^{-1}$) & 0.594 & 0.579 \\
    No. of reflection (unique) & 10964(2699) & 11350(2777) \\
    R$_{int}$ & 0.0386 & 0.579\\
    Completeness to $\theta$ (\%) & 99.0 & 99.4 \\
    Data / restraints / parameters & 2699 / 0 / 199 & 2777 / 0 / 199 \\
    Goodness-of-fit on  $F^{2}$ & 1.143 & 1.106 \\
    Final $R_{1}$ and $wR_{2}$ indices [I $>$ 2$\sigma$(I)] & 0.0524, 0.1411 & 0.0429, 0.1366 \\
    $R_{1}$ and $wR_{2}$ indices (all data) & 0.0536, 0.1684 & 0.0517, 0.1422 \\
    Largest diff. peak and hole (e \mbox{\normalfont\AA}$^{3}$) & 0.272 and -0.225 & 0.172 and -0.168 \\
	
    \hline
  \end{tabular*}
\end{table}

\section{\label{MBD_TS}Comparison between TS and MBD Raman spectra}

\begin{figure}[!ht]
\centering
\includegraphics[scale=0.8]{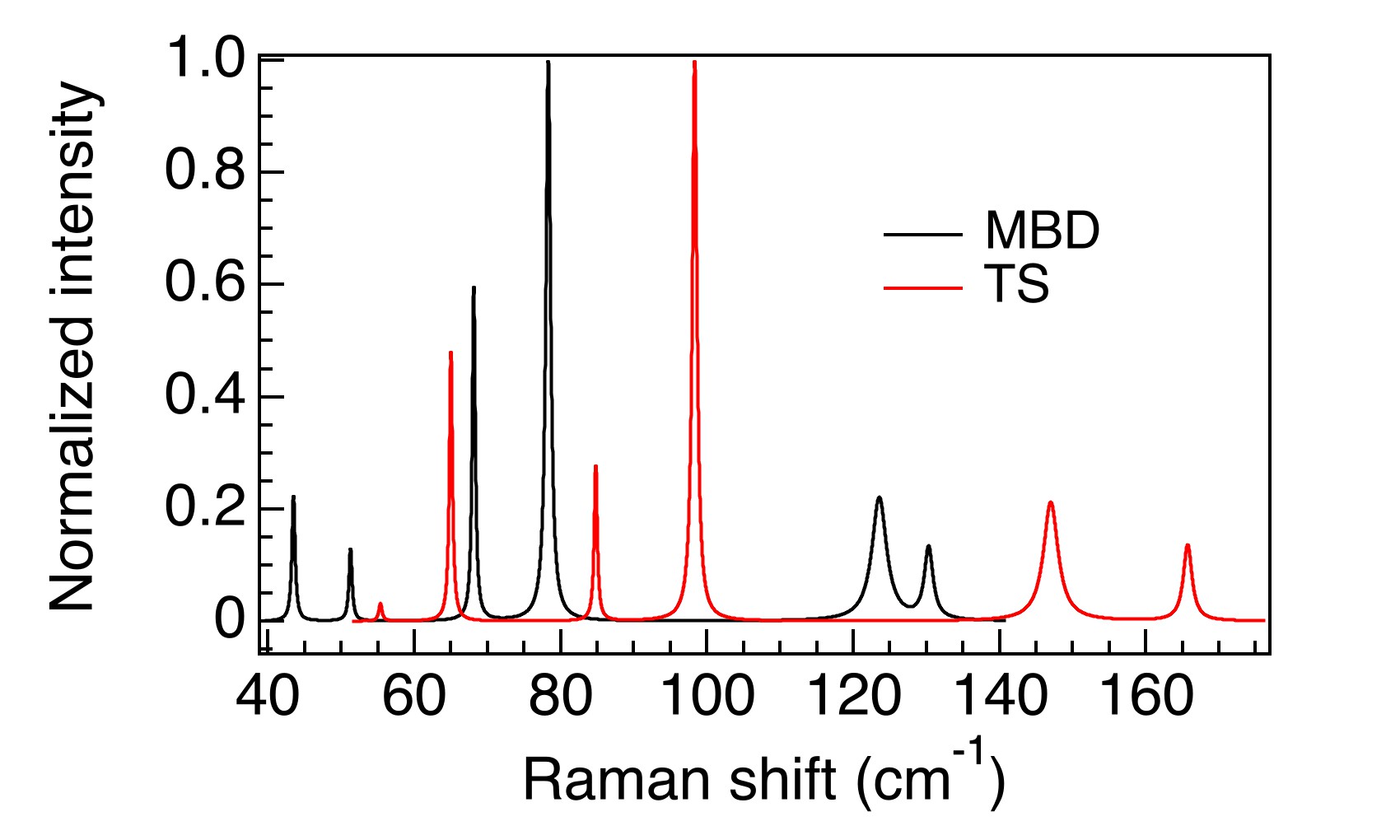}
\caption{\label{fig:anth_MBD_TS}Calculated Raman spectrum of anthracene crystal using the MBD and TS dispersion correction method.}
\end{figure}

\begin{figure}[!ht]
\centering
\includegraphics[scale=0.8]{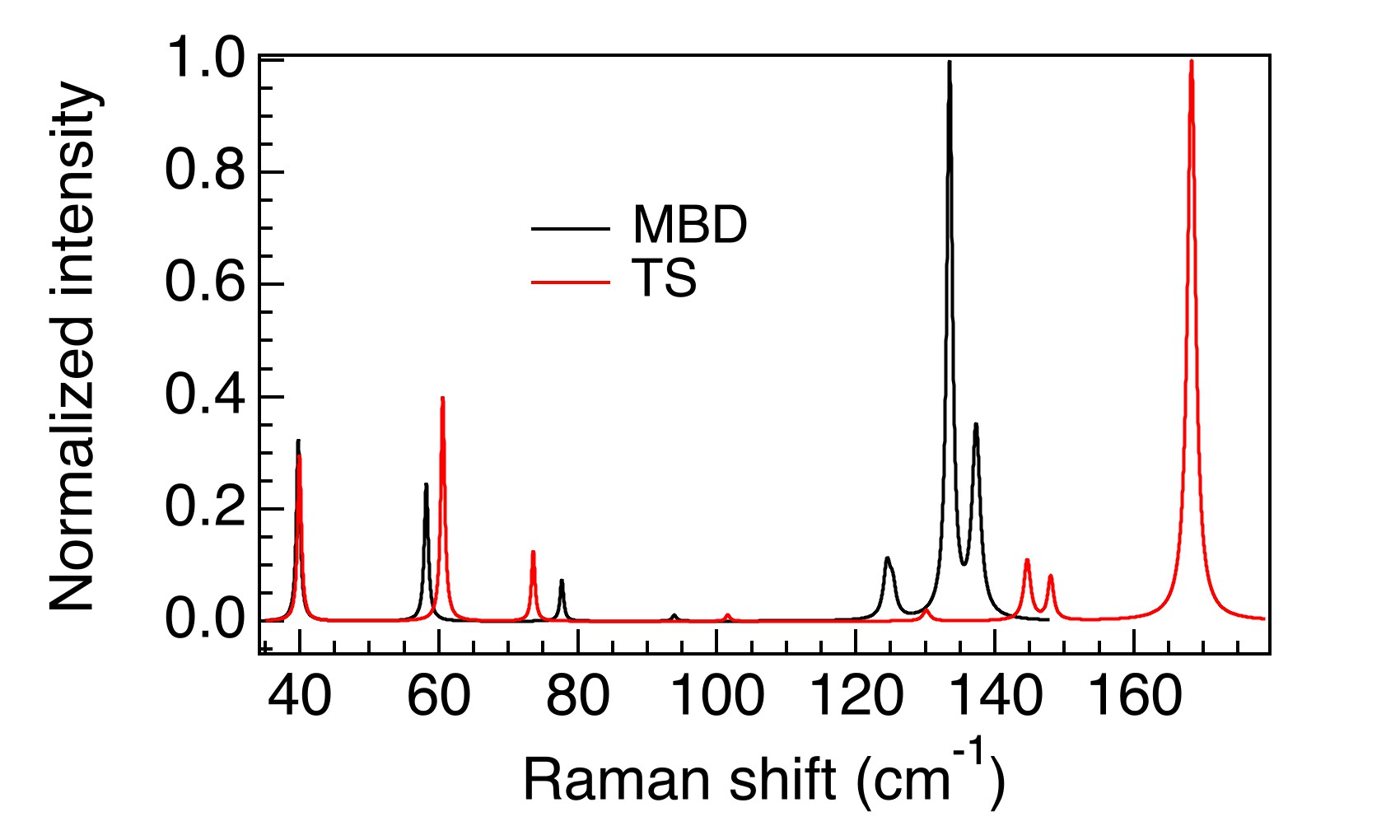}
\caption{\label{fig:pent_MBD_TS}Calculated Raman spectrum of pentacene crystal using the MBD and TS dispersion correction method.}
\end{figure}

Figure \ref{fig:anth_MBD_TS} and Figure \ref{fig:pent_MBD_TS} show a comparison between the unpolarized Raman spectra of anthracene and pentacene crystals, respectively, calculated from DFT with the PBE functional~\cite{Perdew1996} using the Tkatchenko-Scheffler (TS)~\cite{Tkatchenko2009,Kronik2014,Hermann2017,Bedoya-Martinez2018}  and many-body dispersion (MBD)~\cite{Tkatchenko2012,Ambrosetti2014,Bucko2013,Bucko2016} correction method. The results show that the agreement with respect to the experimental data (see main text) is much improved when using the MBD approach.

\section{\label{fit}Raman spectra fitting}

We fit the measured Stokes-shift Raman spectra with the product of the Bose-Einstein distribution and a multi-Lorentzian line shape,

\begin{equation}
I_{Raman}(\omega) = \bigg( \frac{1}{e^{\frac{\hbar \omega}{k_{b}T}    -1}} +1 \bigg) \sum_{i} \frac{A_{i} \Gamma_{i}^{2} \omega _{0,i} \omega  }{\omega ^{2} \Gamma_{i}^{2} +(\omega^{2} -\omega _{0,i}^{2})^{2}}
\label{eq:lorentz}
\end{equation}
Where $\omega_{0,i}$, $A_{i}$ and $\Gamma_{i}$ are the position, intensity and width of each peak respectively, $\omega$ is the measured frequency (Raman shift), $T$ is the temperature, $\hbar$ is the Planck constant and $k_{b}$ is the Boltzmann constant. The Lorentzian in Equation \ref{eq:lorentz} is a variation of the Lorentz oscillator where $A$ is the intensity of the peak.\\

\section{\label{Karnert}Polarization dependent intensity fitting}

We fit the fluctuation of each mode integrated intensity with respect to the incident light polarization according to the model proposed by Kranert et al.\cite{Kranert2016,Kranert2016a} to extract the Raman tensor. We use the following equation,

\begin{equation}
I \propto C(\omega_{p})|e_{i} S \rho J T^{T}Z \mathcal{R} R \mathcal{R}^{-1} T J \rho S^{-1} e_{s}|^{2}\label{eq:Raman_Intensity_Kranert}
\end{equation}
Where $R$ is the Raman tensor of the vibrational mode. The form of the Raman tensors are shown in the main text. $e_{i}$ and $e_{s}$ are the polarization vectors of the incident and scattered light respectively. $e_{i}$ in backscattered configuration and crystal orientation of (001) is,
\begin{equation}
e_{i}=\begin{pmatrix} cos \theta \\
sin \theta \\
0
\end{pmatrix}
\end{equation}
Where $\theta$ in the angle between the polarization vector of the light and an arbitrary axis in the $ab$ plane of the crystal. $e_{s}$ in parallel and perpendicular configurations is,

\begin{equation}
e_{s \parallel}=e_{i} ~,~ e_{s \perp} = e_{i}(\theta + 90^{\circ}) = \begin{pmatrix} -sin \theta \\
cos \theta \\
0
\end{pmatrix}
\end{equation}
$S$ is a rotation matrix, transforming the coordinates of the laboratory system into the coordinate system of the crystals optical axes. According to the crystals orientation and optical properties, we only need to rotated around the $z$ axis as the fast and slow axes are in the $ab$ plane. 

$J$ is the Jones matrix accounting for the birefringence effect. The form of the Jones matrix is,

\begin{equation}
J =\begin{pmatrix} 1 & 0 & 0 \\
0 & e^{i \delta} & 0\\
0 & 0 & 0
\end{pmatrix}
\end{equation}
Where $\delta$ is the phase shift between the light with polarization parallel to the fast and the slow axis.

$\rho$ corrects for the different reflection coefficients due to the birefringent nature of the crystals. Where $\rho$ is,
\begin{equation}
\rho =\begin{pmatrix} \frac{1-r_{fast}}{1-r_{slow}} & 0 & 0 \\
0 & 1 & 0\\
0 & 0 & 0
\end{pmatrix}
\end{equation}
$r_{fast}$ and $r_{slow}$ are the reflection coefficients in the fast and slow axis respectively which are defined as,
\begin{equation}
r_{fast,slow} = \frac{n_{fast,slow}-1}{n_{fast,slow}+1}
\end{equation}
where $n_{fast}$ and $n_{slow}$ are the refractive indices in the fast and slow axis respectively. For anthracene\cite{Nakada1962},
\begin{equation}
n_{fast} =1.55 ~ ,~ n_{slow} = 1.78
\end{equation}
and for pentacene\cite{Faltermeier2006},

\begin{equation}
n_{fast} = 1.28~ ,~ n_{slow} = 1.85
\end{equation}
$Z$ corrects for the effect of the refractive index on the light collection angle. Its form is,

\begin{equation}
Z =\begin{pmatrix} \sqrt{\frac{\Omega _{slow}}{\Omega _{fast}}} & 0 & 0 \\
0 & 1 & 0\\
0 & 0 & 0
\end{pmatrix}
\end{equation}
Where,
\begin{equation}
\Omega_{fast,slow} = 4 \pi sin^{2} \bigg( \frac{arcsin\frac{NA}{n_{fast,slow}}}{4} \bigg)
\end{equation}
Where NA is the numerical aperture of the objective used in the system which is 0.55.

$C(\omega_{p})$ corrects for the dependence of the intensity on the phonon frequency $\omega_{p}$ and incident photon frequency and its form is,
\begin{equation}
C(\omega_{p}) = \frac{\omega_{i}(\omega_{i}-\omega_{p})^{3}}{\omega_{p}}
\end{equation}
In our case we use a 785~nm laser and the phonon frequencies were extracted from the Raman spectra fit (Equation \ref{eq:lorentz}).
$T$ is the transformation matrix transforming the external to the allowed internal polarizations, $\mathcal{R}$ is a rotational matrix rotating the Raman tensor from the principle axes coordinates to the optical coordinates. We found both to be irrelevant to our analysis.\\

\section{\label{global}Global fitting}

In the global fitting procedure, we fit the polarization dependence of all the peaks in both parallel and perpendicular configurations according to Equation \ref{eq:Raman_Intensity_Kranert}. We fix the zero components of the Raman tensors of anthracene. We link the Raman tensors for parallel and perpendicular configurations for each peak (the fit had to produce the same Raman tensor for parallel and perpendicular configurations for each mode). We also link $S$ and $\delta$ to all modes in both configurations (for the entire measurement).

Table \ref{tbl:Raman_tensor_anthracene}, Table \ref{tbl:Raman_tensor_anthracene_DFT},  Table \ref{tbl:Raman_tensor_pentacene}  and Table \ref{tbl:Raman_tensor_pentacene_DFT} show the Raman tensors we obtain for anthracene and pentacene at 10~K and from DFT calculations. Since in the model for the Raman intensity the Raman tensor is squared, we have no access to the sign of the Raman tensor components. Also, since the experiment is on the (001) crystal orientation we only have access to the $xy$ components of the Raman tensors. We normalize the values of the Raman tensors to the highest absolute value among all tensors. The values we obtain for $\delta$ are 45.6$^{\circ}$ and 90.1$^{\circ}$ for anthracene and pentacene respectively. For the $A_{g}^{(5)}$ mode of pentacene, the DFT result show two very close Raman active modes (124.8~$cm^{-1}$ and 125.2~$cm^{-1}$). We assume that they are degenerate experimentally and combine the intensities from both to represent the polarization-dependent signal of $A_{g}^{(5)}$.\\

\begin{table}[ht]
\large
  \caption{Raman tensors of anthracene obtained from global fitting the polarization-orientation Raman measurement at 10~K}
\center
  \label{tbl:Raman_tensor_anthracene}
  \begin{tabular*}{0.7\textwidth}{@{\extracolsep{\fill}}lllllll}
    \hline
    Component & $A_{g}^{(1)}$ & $B_{g}^{(1)}$ & $B_{g}^{(2)}$ & $A_{g}^{(2)}$ & $A_{g}^{(3)}$ & $B_{g}^{(3)}$ \\
    \hline
    a & 0.10 & 0.00 & - & 0.23 & 0.41 & 0.00 \\
    b & 0.34 & 0.00 & - & 0.26 & 1.00 & 0.00 \\
    d & 0.00 & 0.11 & - & 0.00 & 0.00 & 0.44 \\
    \hline
  \end{tabular*}
\end{table}

\begin{table}[ht]
\large
  \caption{Raman tensors of anthracene obtained from DFT}
  \center
  \label{tbl:Raman_tensor_anthracene_DFT}
  \begin{tabular*}{0.7\textwidth}{@{\extracolsep{\fill}}lllllll}
    \hline
    Component & $A_{g}^{(1)}$ & $B_{g}^{(1)}$ & $B_{g}^{(2)}$ & $A_{g}^{(2)}$ & $A_{g}^{(3)}$ & $B_{g}^{(3)}$ \\
    \hline
    a & 0.20 & 0.00 & 0.00 & -1.00 & 0.07 & 0.00 \\
    b & -0.19 & 0.00 & 0.00 & 0.20 & 0.93 & 0.00 \\
    d & 0.00 & -0.16 & -0.39 & 0.00 & 0.00 & 0.42 \\
    \hline
  \end{tabular*}
\end{table}

\begin{table}[ht]
\large
  \caption{Raman tensors of pentacene obtained from global fitting the polarization-orientation Raman measurement at 10~K}
  \center
  \label{tbl:Raman_tensor_pentacene}
  \begin{tabular*}{0.7\textwidth}{@{\extracolsep{\fill}}llllllll}
    \hline
    Component & $A_{g}^{(1)}$ & $A_{g}^{(2)}$ & $A_{g}^{(3)}$ & $A_{g}^{(4)}$ & $A_{g}^{(5)}$ & $A_{g}^{(6)}$ & $A_{g}^{(7)}$ \\
    \hline
    a & 0.28 & 0.26 & 0.18 & - & 1.00 & 0.36 & 0.48\\
    b & 0.01 & 0.02 & 0.01 & - & 0.02 & 0.09 & 0.19 \\
    d & 0.06 & 0.09 & 0.05 & - & 0.18 & 0.19 & 0.53 \\
    \hline
  \end{tabular*}
\end{table}

\begin{table}[ht]
\large
  \caption{Raman tensors of pentacene obtained from DFT}
  \center
  \label{tbl:Raman_tensor_pentacene_DFT}
  \begin{tabular*}{0.8\textwidth}{@{\extracolsep{\fill}}lllllllll}
    \hline
    Component & $A_{g}^{(1)}$ & $A_{g}^{(2)}$ & $A_{g}^{(3)}$ & $A_{g}^{(4)}$ & \multicolumn{2}{c}{$A_{g}^{(5)}$}  & $A_{g}^{(6)}$ & $A_{g}^{(7)}$\\
    \hline
    a & 0.00 & -0.29 & -0.16 & -0.12 & 0.52 & 0.34 & -0.05 & 0.25 \\
    b & -0.27 & -0.12 & -0.16 & 0.01 & -0.08 & -0.13 &  0.24 & 0.18 \\
    d & -0.13 & -0.19 & -0.02 & 0.00 & -0.04 & 0.10 &  -1.00 & -0.70 \\
    \hline
  \end{tabular*}
\end{table}

\section{\label{temp_PO_all}Polarization-orientation Raman measurements at different temperatures}
\begin{figure}[!ht]
\centering
\includegraphics[scale=0.11]{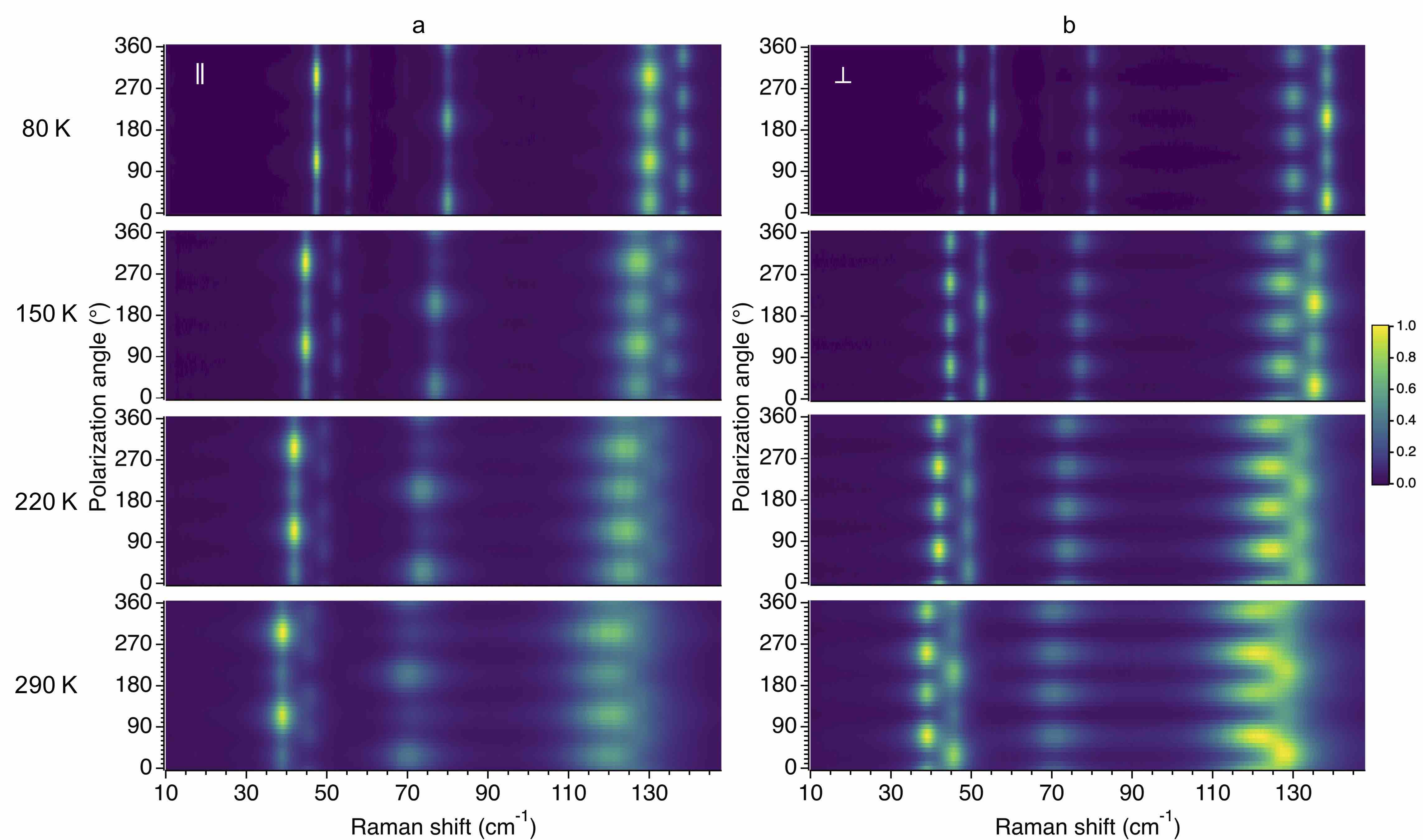}
\caption{\label{fig:anth_T_PO_image}The polarization-orientation Raman measurements of anthracene in (a) parallel and (b) perpendicular configurations at 80~K, 150~K, 220~K and 290~K.}
\end{figure}

\begin{figure}[!ht]
\centering
\includegraphics[scale=0.11]{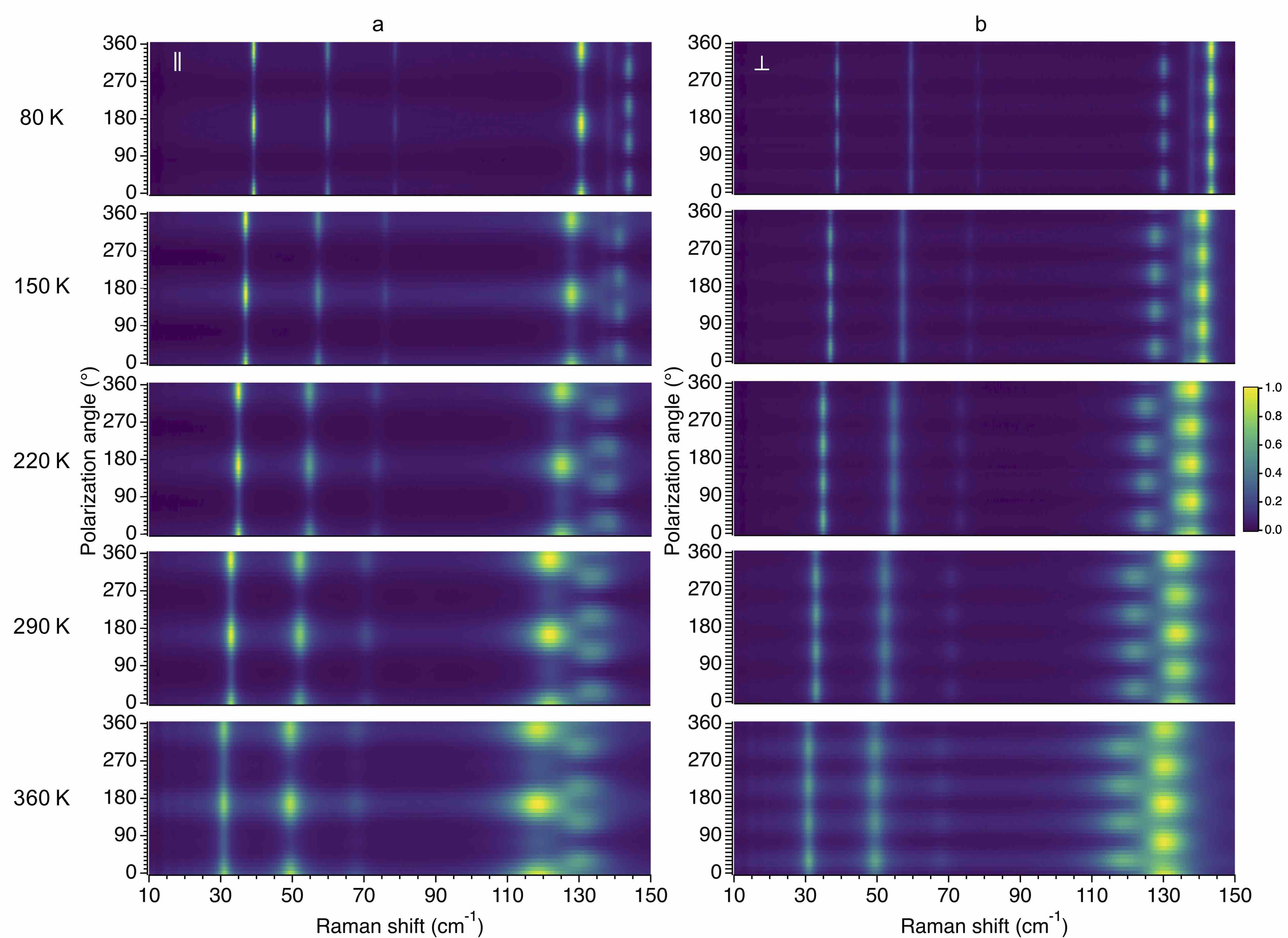}
\caption{\label{fig:pent_T_PO_image}The polarization-orientation Raman measurements of pentacene in (a) parallel and (b) perpendicular configurations at 80~K, 150~K, 220~K, 290~K and 360~K.}
\end{figure}

Figure \ref{fig:anth_T_PO_image} and Figure \ref{fig:pent_T_PO_image} present the raw data we obtain from the polarization-orientation Raman measurements for both parallel and perpendicular configurations for single crystals of anthracene and pentacene with crystal orientation of (001) at different temperatures.

\section{\label{temp_PO_crs}Temperature dependent polarization-orientation Raman - perpendicular configuration}

\begin{figure}[!ht]
\centering
\includegraphics[scale=0.15]{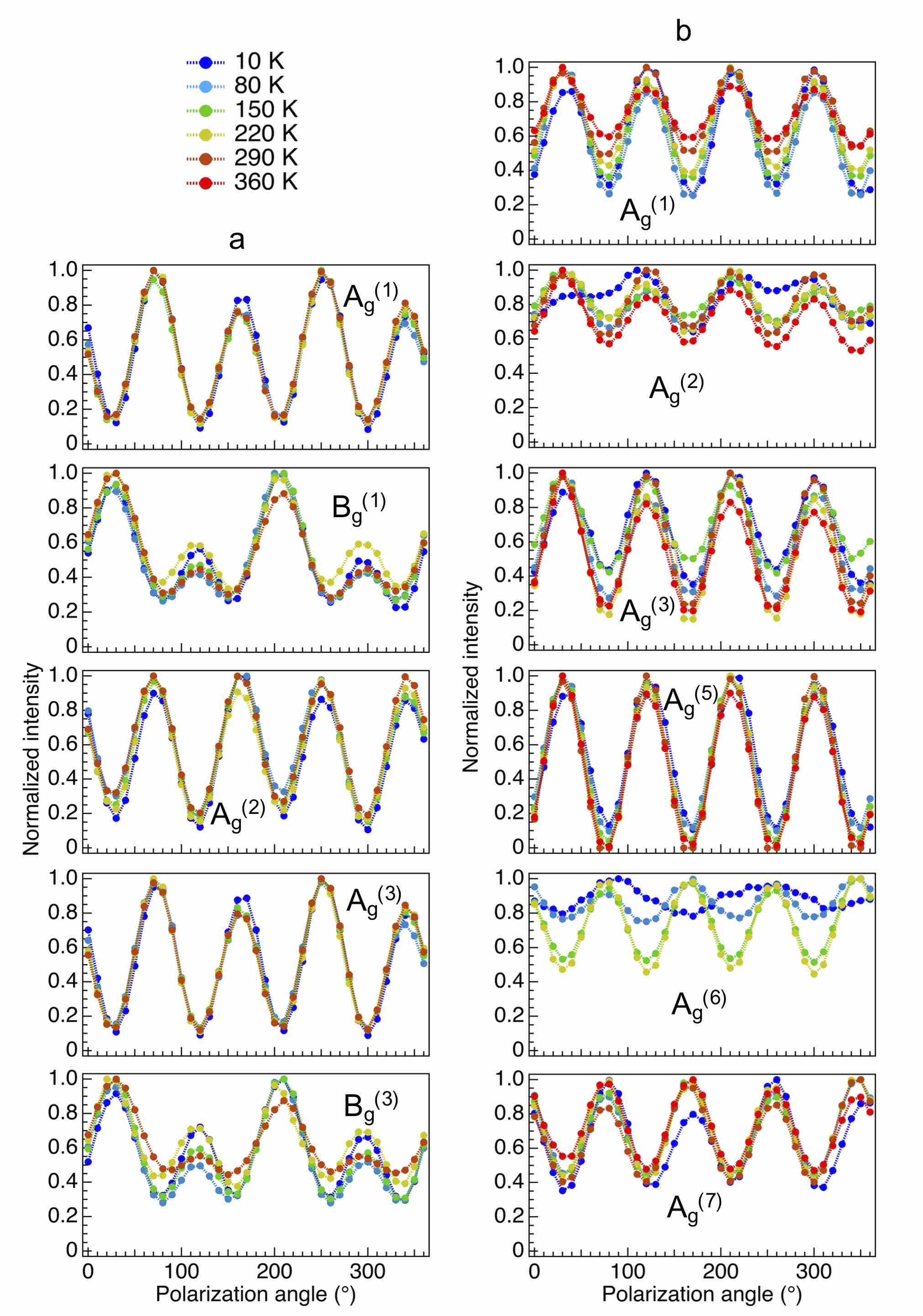}
\caption{\label{fig:PO_temp_crs}Experimental temperature-dependence of the polarization-dependence for the six lowest-frequency modes of (a) anthracene and (b) pentacene in perpendicular configuration. The intensities were normalized for each mode at each temperature.}
\end{figure}

Figure \ref{fig:PO_temp_crs} presents the intensity fluctuations for each peak with respect to the polarization angle in the perpendicular configuration for anthracene and pentacene at different temperatures.\\

\section{\label{liquid}Polarization dependent measurement of a liquid}

\begin{figure}[!ht]
\centering
\includegraphics[scale=0.15]{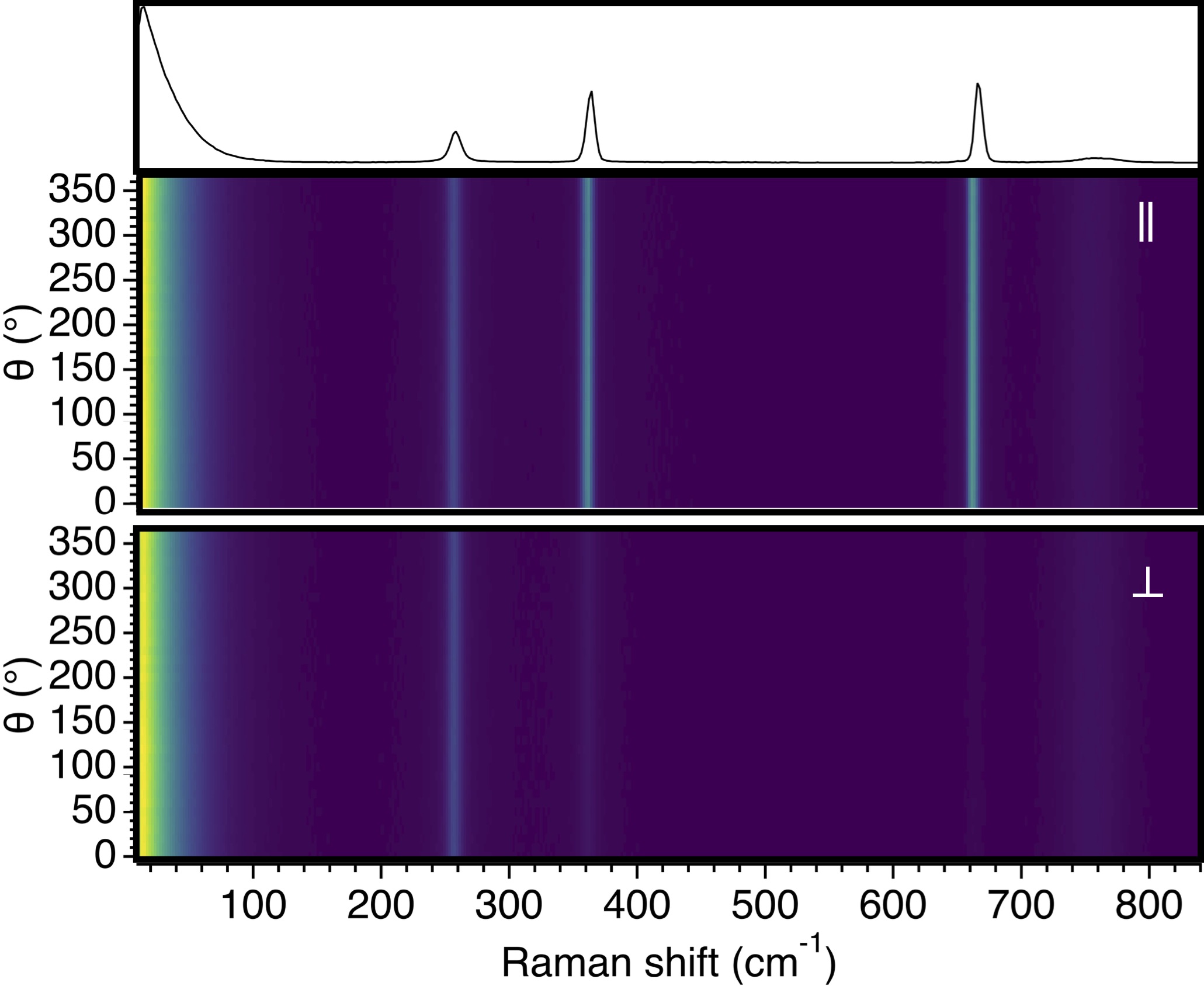}
\caption{\label{fig:PO_chloroform}Polarization dependent measurement of liquid chloroform in perpendicular and parallel configurations. The top part of the figure shows a typical spectrum in the parallel configuration.}
\end{figure}

Figure \ref{fig:PO_chloroform} presents the polarization-dependent measurement of chloroform (CHCl$_{3}$) in parallel and perpendicular configurations, showing the measured Raman spectrum is independent of the polarization angle of the incident light.\\

\section{\label{T_dependent}Temperature dependent Raman spectroscopy of anthracene and pentacene}

\begin{figure}[!ht]
\centering
\includegraphics[scale=0.18]{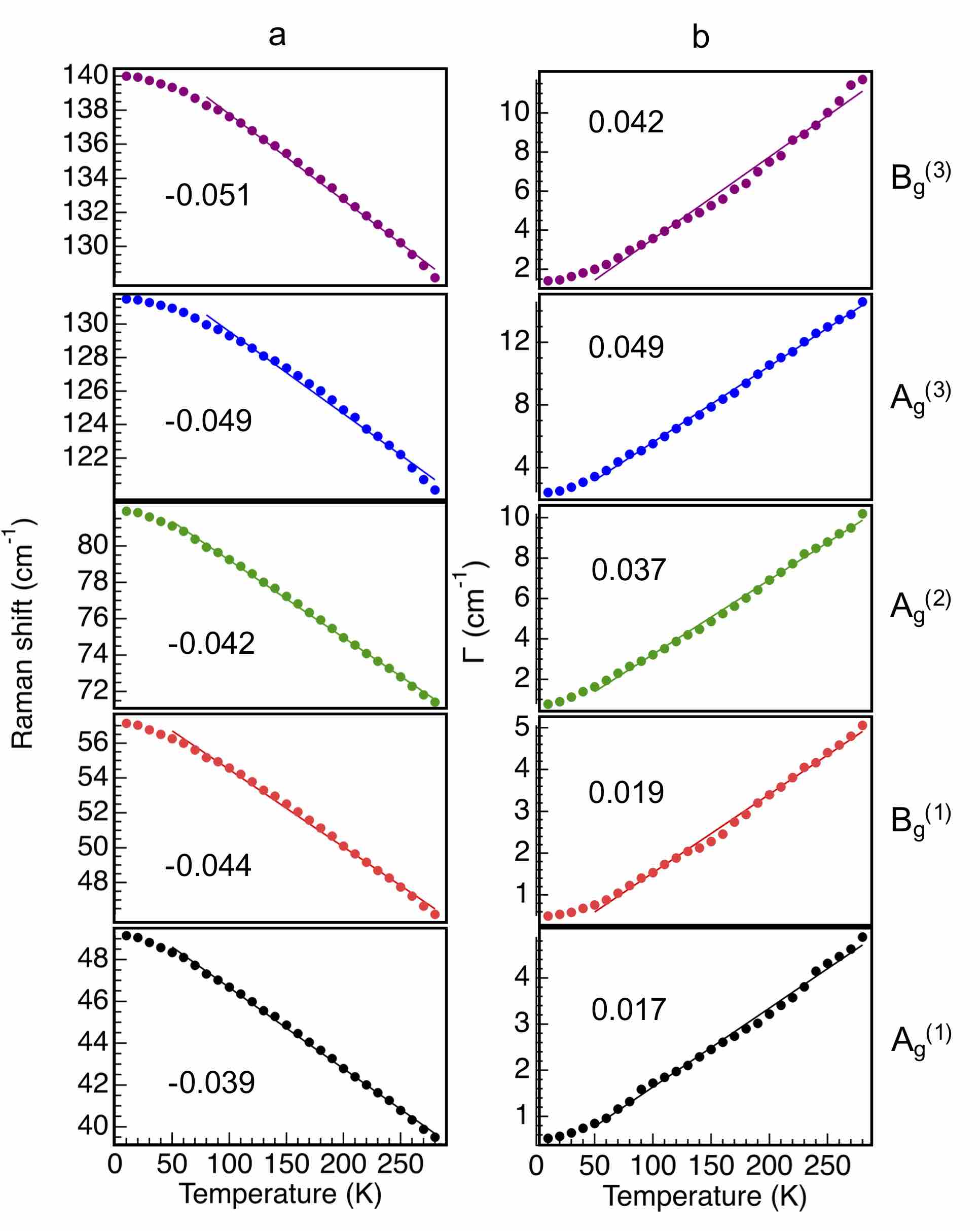}
\caption{\label{fig:ant_T_dep}The temperature dependence of the peaks (a) position and (b) width of anthracene. The number in each plot represents the slope of the graph. The dots represent the data and the solid lines represent the fit to a linear line.}
\end{figure}

\begin{figure}[!ht]
\centering
\includegraphics[scale=0.18]{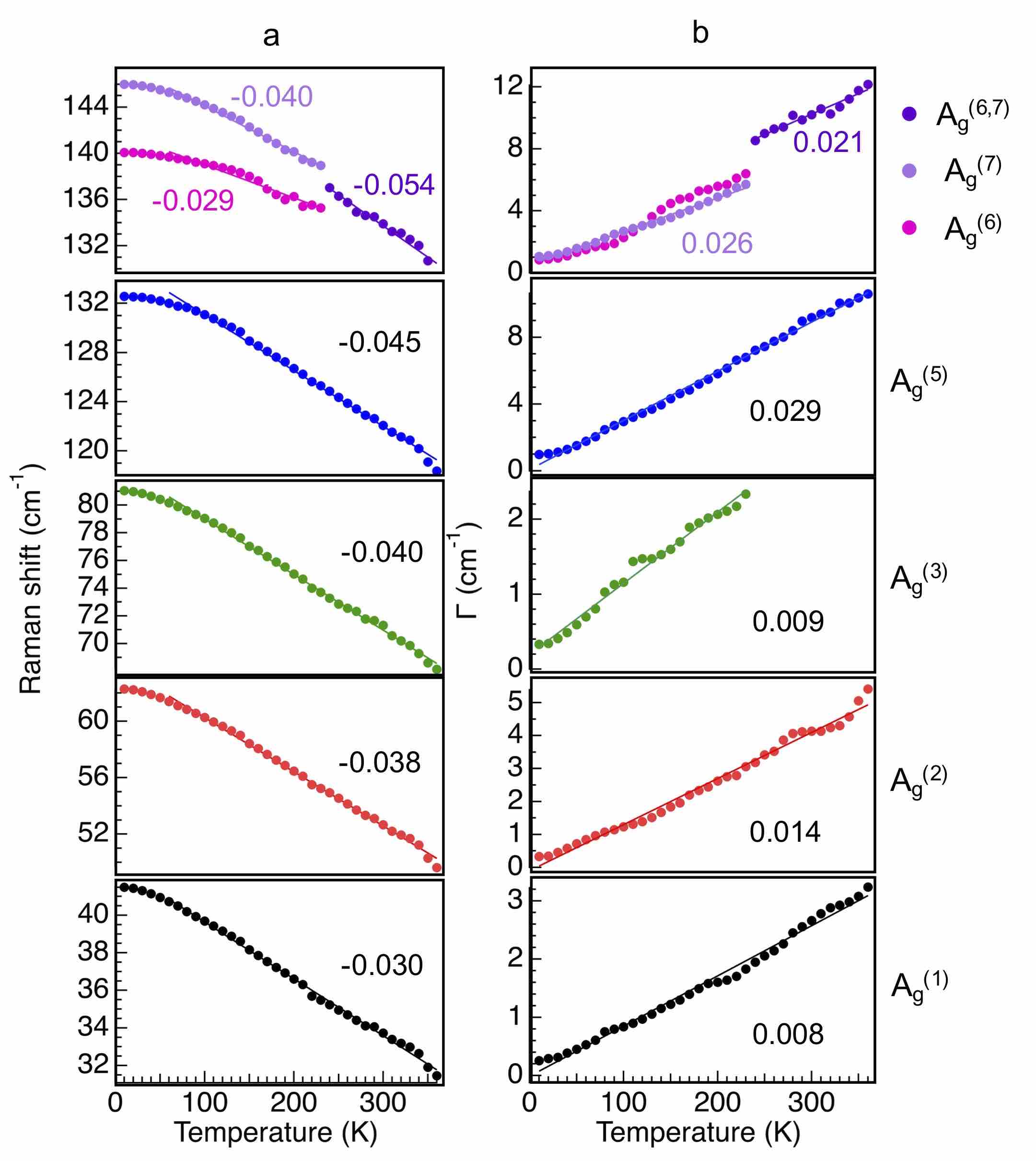}
\caption{\label{fig:pent_T_dep}The temperature dependence of the peaks (a) position and (b) width of pentacene. The number in each plot represents the slope of the graph. The dots represent the data and the solid lines represent the fit to a linear line.}
\end{figure}

Figure \ref{fig:ant_T_dep} and Figure \ref{fig:pent_T_dep} show the temperature dependence of the peaks position and width of anthracene and pentacene. We extract the slope of the graphs by fitting the data to a linear line. We exclude the data at lower temperatures from the fitting process, where the dependency is not linear. For pentacene, from 240~C we can not deconvolve $A_{g}^{(6)}$ and $A_{g}^{(7)}$, thus we fit it as one peak labelled $A_{g}^{(6,7)}$. For $A_{g}^{(3)}$, from 240~C the peak intensity is low (close to the noise level) so the peak width we extract is less reliable.

\begin{figure}[!ht]
\centering
\includegraphics[scale=0.12]{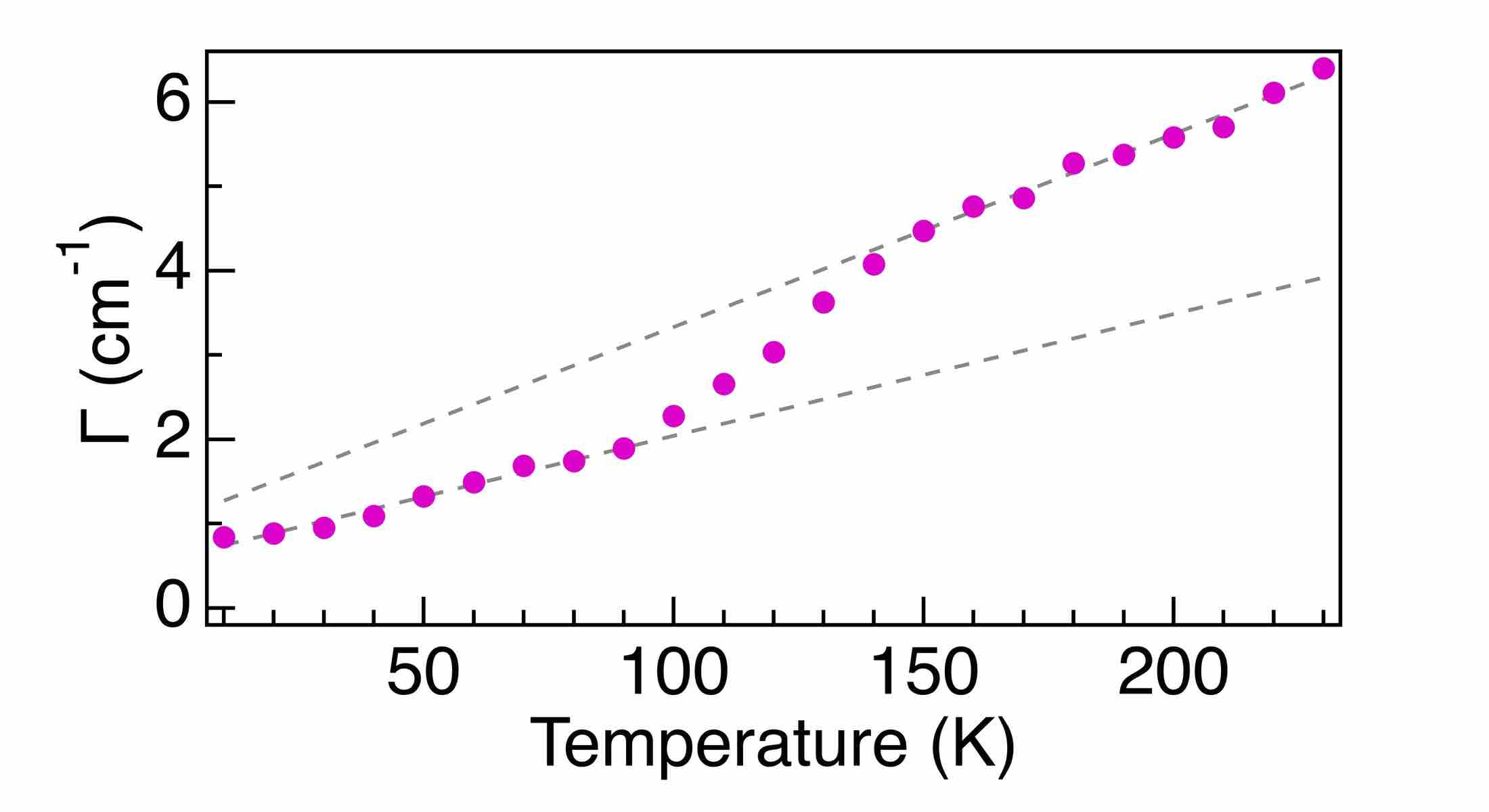}
\caption{\label{fig:pent_w5_width}The temperature dependence of $A_{g}^{(6)}$ width of pentacene. The dashed lines show the two linear regions on the two polymorphs.}
\end{figure}

Figure \ref{fig:pent_w5_width} shows a zoom-in on the temperature dependence of pentacene $A_{g}^{(6)}$ width. Though usually, the temperature dependence of a peak width is linear, here we can see two different regions indicating a phase transition between two different polymorphs. The temperature range of this transition coincides with the range of the phase transition shown by temperature-dependent polarization-orientation Raman in the main text.\\

\section{\label{DOS}Phonon density of states}
\begin{figure}[!ht]
\centering
\includegraphics[scale=0.6]{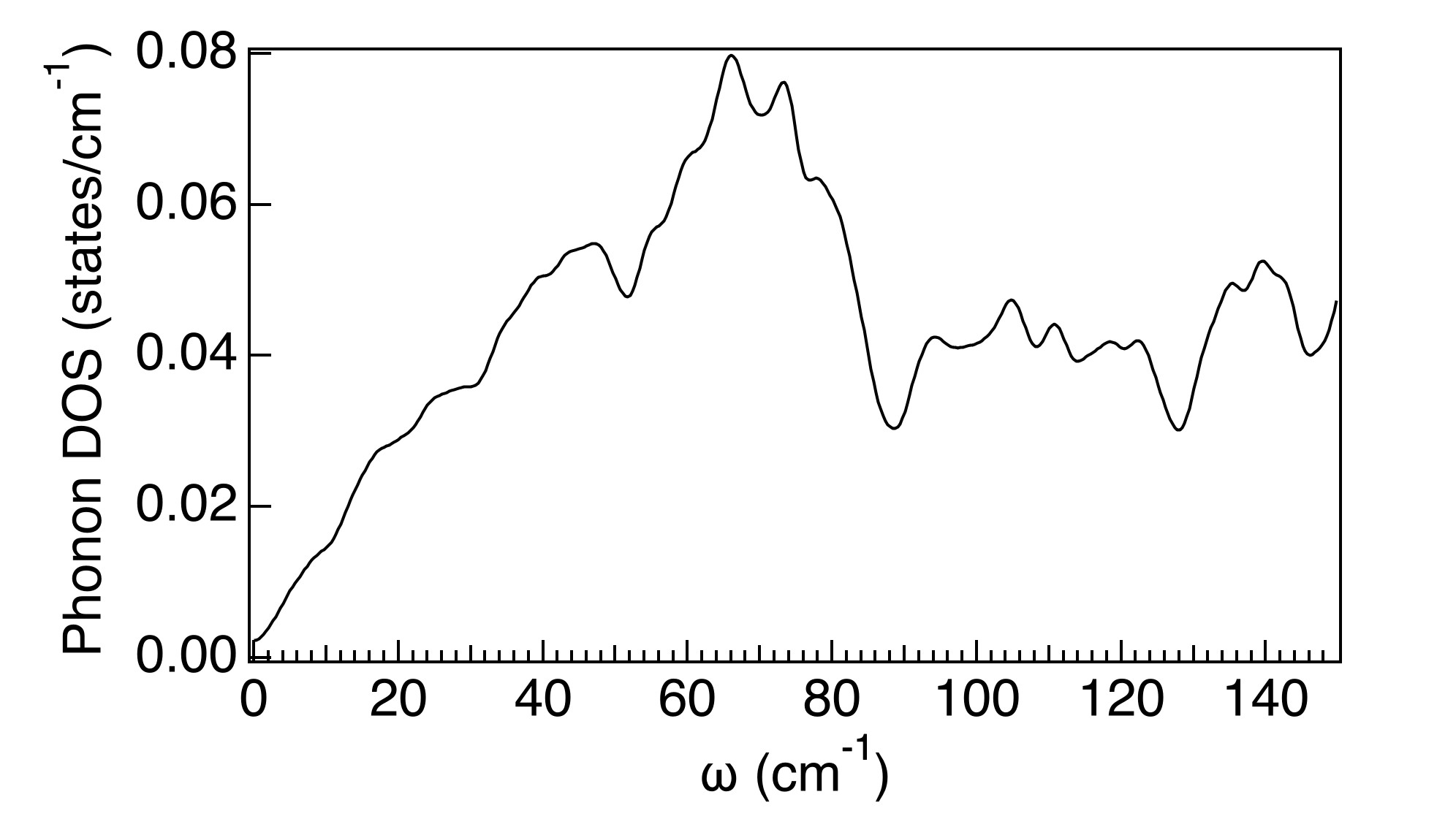}
\caption{\label{fig:DOS_ant}DFT-calculated phonon density of states of anthracene.}
\end{figure}

\begin{figure}[!ht]
\centering
\includegraphics[scale=0.6]{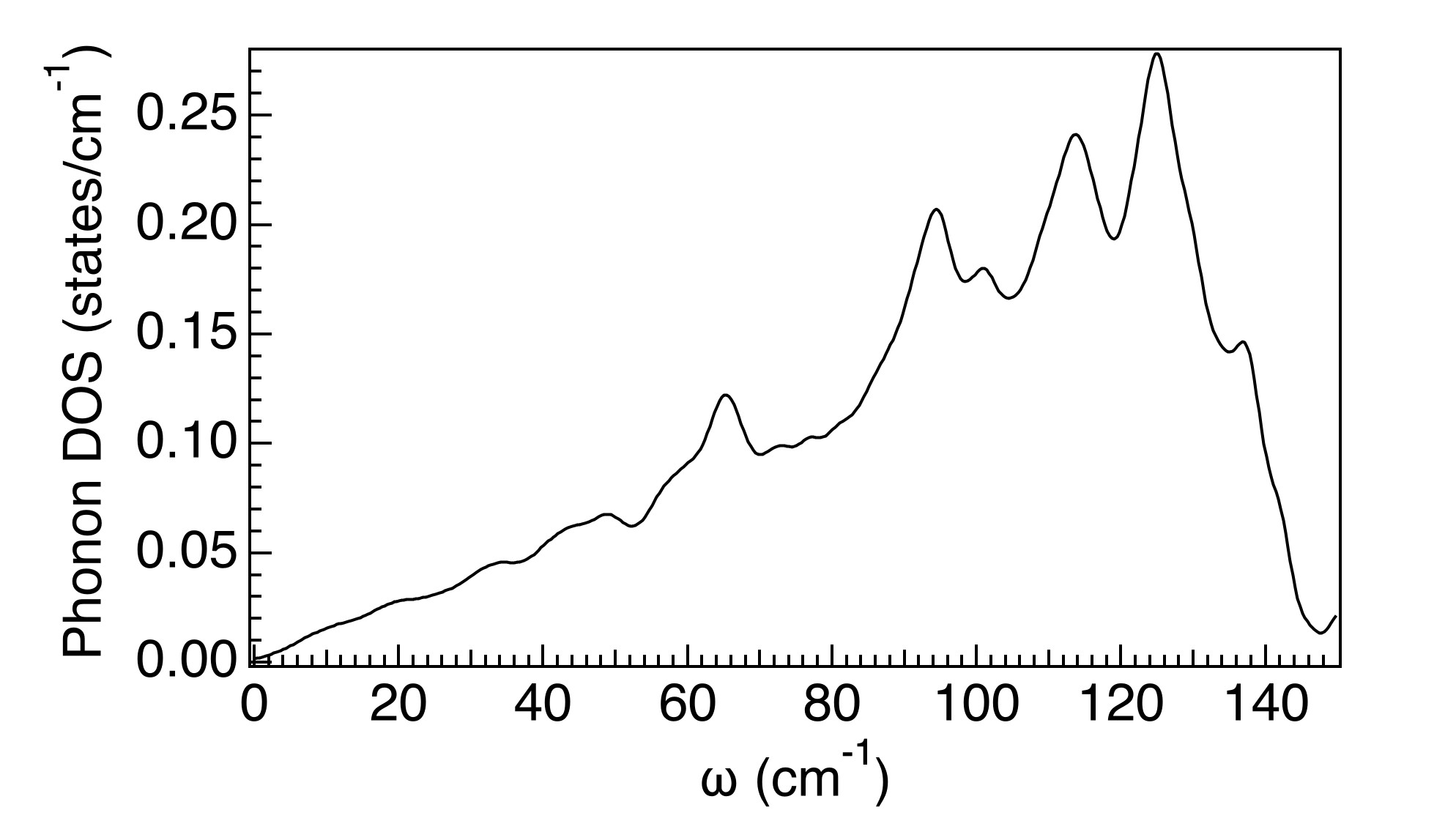}
\caption{\label{fig:DOS_pent}DFT-calculated phonon density of states of pentacene.}
\end{figure}

Figure \ref{fig:DOS_ant} and \ref{fig:DOS_pent} present the DFT-calculated phonon density of states (phonon DOS) of anthracene and pentacene.
The phonon DOS depending on the frequency $ \omega $ is defined as
\begin{equation} \label{eq:phonon DOS}
	D(\omega) = \frac{1}{N} \sum_{\vec{q},i} \delta\left( \omega-\omega(\vec{q}_{i}) \right),
\end{equation}
where $ \delta $ denotes the Dirac delta function and $ \omega(\vec{q}_{i}) $ are the phonon frequencies on a sampling mesh of $ \vec{q} $-points in the first Brillouin zone (BZ) with a number of $ N $ grid points. Here, the phonon DOS is computed with the phonopy package on a discretized set of frequencies $ \omega $ in the range up to $ \omega = 150~cm^{-1} $, incorporating all the low-frequency phonon modes. To consider line width broadening, the delta function in Equation~\ref{eq:phonon DOS} is replaced by a Gaussian distribution with a broadening of $ 1.33~cm^{-1} $. For sampling the BZ, a $ 30 \times 30 \times 30 $ $ \vec{q} $-points grid was applied.

\section{\label{Raman}Polarization optical components scheme}

\begin{figure}[!ht]
\centering
\includegraphics[scale=0.15]{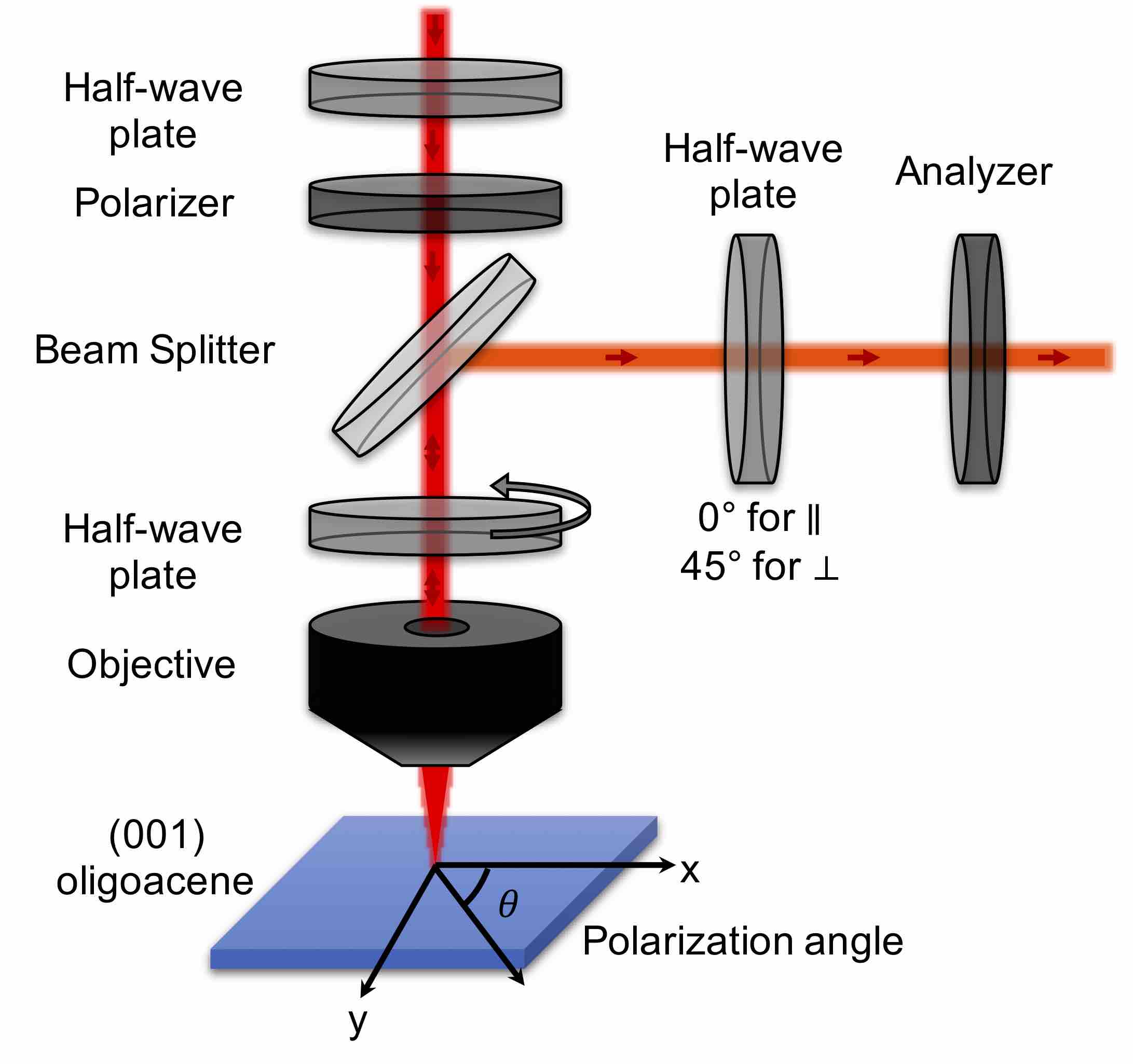}
\caption{\label{fig:PO_setup}Scheme of the polarization optical components experimental set-up.}
\end{figure}

In this study, we use our state-of-the-art home-built Raman system as described in the main text.
Figure \ref{fig:PO_setup} shows a scheme of the system, focusing on the optical components which control the light polarization. This scheme allows the use of only half-wave plates to control the angle of light polarization during the measurement, avoiding moving the polarizers which cause beam deviation.\\ 

\bibliographystyle{abbrv}

\renewcommand{\thepage}{S\arabic{page}}  
\renewcommand{\thesection}{S\arabic{section}}   
\renewcommand{\thetable}{S\arabic{table}}   
\renewcommand{\thefigure}{S\arabic{figure}}